\documentclass[prd,reprint,nofootinbib,amsmath,amssymb,aps,floatfix]{revtex4-1}
\usepackage{bm}
\usepackage{amssymb,amsmath,graphicx}
\usepackage[linktoc=page]{hyperref}
\usepackage{subfigure}
\usepackage{epsfig}
\usepackage{amsfonts}
\usepackage{color}
\definecolor{arvin}{RGB}{250,0,250}
\usepackage[normalem]{ulem}
\usepackage{dsfont}
\usepackage{braket}
\DeclareMathOperator{\Tr}{Tr}

\newcommand{\mR}{\mathcal{R}}

\newcommand{\cse}{\mathcal{C} \subseteq \mathcal{E}}

\newcommand{\be}{\begin{equation}}
\newcommand{\ee}{\end{equation}}

\begin{document}
\title{Local Modular Hamiltonians from the Quantum Null Energy Condition}
\author{Jason Koeller}
\email{jkoeller@berkeley.edu}
\author{Stefan Leichenauer}
\email{sleichen@berkeley.edu}
\author{Adam Levine}
\email{arlevine@berkeley.edu}
\author{Arvin Shahbazi Moghaddam}
\email{arvinshm@berkeley.edu}
\affiliation{Center for Theoretical Physics and Department of Physics\\
University of California, Berkeley, CA 94720, USA 
}%
\affiliation{Lawrence Berkeley National Laboratory, Berkeley, CA 94720, USA}
\begin{abstract}
The vacuum modular Hamiltonian \(K\) of the Rindler wedge in any relativistic quantum field theory is given by the boost generator. Here we investigate the modular Hamiltoninan for more general half-spaces which are bounded by an arbitrary smooth cut of a null plane. We derive a formula for the second derivative of the modular Hamiltonian with respect to the coordinates of the cut which schematically reads \(K'' = T_{vv}\). This formula can be integrated twice to obtain a simple expression for the modular Hamiltonian. The result naturally generalizes the standard expression for the Rindler modular Hamiltonian to this larger class of regions. Our primary assumptions are the quantum null energy condition --- an inequality between the second derivative of the von Neumann entropy of a region and the stress tensor --- and its saturation in the vacuum for these regions. We discuss the validity of these assumptions in free theories and holographic theories to all orders in \(1/N\). 
\end{abstract}
\maketitle


\section{Introduction and Summary}
\label{sec-intro}
The reduced density operator \(\rho\) for a region in quantum field theory encodes all of the information about observables localized to that region. Given any \(\rho\), one can define the \emph{modular Hamiltonian} \(K\) by
\begin{align}\label{Kdef}
	\rho = e^{-K}.
\end{align}
Knowledge of this operator is equivalent to knowledge of $\rho$, but the modular Hamiltonian frequently appears in calculations involving entanglement entropy. In general, i.e. for arbitrary states reduced to arbitrary regions, \(K\) is a complicated non-local operator. However, in certain cases it is known to simplify.

The most basic example where $K$ simplifies is the vacuum state of a QFT in Rindler space, i.e. the half-space \(t=0, x \geq 0\). The Bisognano--Wichmann theorem \cite{Bisognano:1976za} states that in this case the modular Hamiltonian is
\begin{align}\label{KRindler}
	\Delta K = \frac{2\pi}{\hbar}\int d^{d-2}y \int_{0}^{\infty} x \, T_{tt} \, dx
\end{align}
where \(\Delta K \equiv K - \braket{K}_{\rm vac}\) defines the vacuum-subtracted modular Hamiltonian, and \(y\) are \(d-2\) coordinates parametrizing the transverse directions. The vacuum subtraction generally removes regulator-dependent UV-divergences in \(K\). Other cases where the modular Hamiltonian is known to simplify to an integral of local operators are obtained via conformal transformation of Eq.~\eqref{KRindler}, including spherical regions in CFTs \cite{Casini:2011kv}, regions in a thermal state of 1+1 CFTs \cite{Cardy:2016fqc}, and null slabs \cite{Bousso:2014sda,Bousso:2014aa}.

\begin{figure}[]\label{fig-region}
	\includegraphics[width=.35\textwidth]{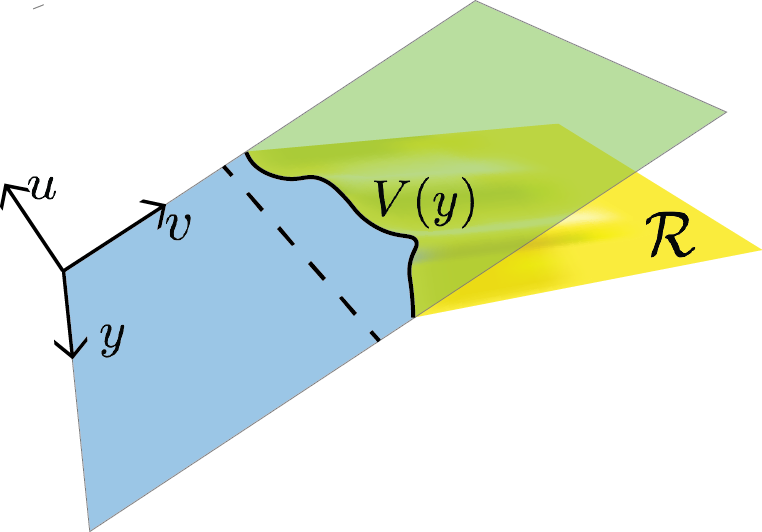}
	\caption{This image depicts a section of the plane $u = t-x =0$. The region $\mathcal{R}$ is defined to be one side of a Cauchy surface split by the codimension-two entangling surface $\partial \mathcal{R} = \lbrace{ (u = 0, v = V(y), y) \rbrace}$. The dashed line corresponds to a flat cut of the null plane.}
\end{figure}

Using conservation of the energy-momentum tensor, one can easily re-express the Rindler modular Hamiltonian in Eq.~\eqref{KRindler} as an integral over the future Rindler horizon \(u\equiv t-x =0\) which bounds the future of the Rindler wedge:
\begin{align}\label{KRindlerNull}
	\Delta K = \frac{2\pi}{\hbar} \int d^{d-2}y \int_{0}^{\infty} v \,T_{vv} \, dv,
\end{align}
where \(v \equiv t+x \). It is important to note that standard derivations of \eqref{KRindler} or \eqref{KRindlerNull}, e.g. \cite{Bisognano:1976za,Casini:2011kv}, do not apply when the entangling surface is defined by a non-constant cut of the Rindler horizon (see Fig.~\ref{fig-region}). One of the primary goals of this paper is to provide such a derivation. 

For a large class of quantum field theories satisfying a precise condition specified momentarily, we will show that the vacuum modular Hamiltonian for the region \(\mR[V(y)]\) above an arbitrary cut \(v = V(y)\) of a null plane is given by
\begin{align}\label{KWigglyNull}
	\Delta K = \frac{2\pi}{\hbar} \int d^{d-2} y \int_{V(y)}^{\infty} (v - V(y)) \,T_{vv} \, dv
\end{align}
This equation has been previously derived by Wall for free field theories \cite{Wall:2011hj} building on \cite{Burkardt:1995ct,Sewell:1982zz}, and to linear order in the deformation away from \(V(y) = \rm const\) in general QFTs by Faulkner et al. \cite{Faulkner:2016mzt}. In CFTs, conformal transformations of Eq.~\eqref{KWigglyNull} yield versions of the modular Hamiltonian for non-constant cuts of the causal diamond of a sphere. 

The condition leading to Eq.~\eqref{KWigglyNull} is that the theory should satisfy the \emph{quantum null energy condition} (QNEC) \cite{Bousso:2015mna,Bousso:2015wca,Koeller:2015qmn,Akers:2016ugt} --- an inequality between the stress tensor and the von Neumann entropy of a region --- and saturate the QNEC in the vacuum for regions defined by cuts of a null plane. We will review the statement of the QNEC in Sec.~\ref{sec:main}.

The QNEC has been proven for free and superrenormalizable \cite{Bousso:2015wca}, as well as holographic \cite{Koeller:2015qmn, Akers:2016ugt} quantum field theories. We take this as reasonable evidence that the QNEC is a true fact about relativistic quantum field theories in general, and for the purposes of this paper take it as an assumption. In Sec.~\ref{sec:main} we will show how saturation of the QNEC in a given state leads to an operator equality relating certain derivatives of the modular Hamiltonian of that state to the energy-momentum tensor. Applied to the case outlined above, this operator equality will be integrated to give Eq.~\eqref{KWigglyNull}.

Given the argument in Sec.~\ref{sec:main}, the only remaining question is whether the QNEC is in fact saturated in the vacuum state for entangling surfaces which are cuts of a null plane. This has been shown for free theories in \cite{Bousso:2015wca}. In Sec.~\ref{sec-calculation}, we prove that this is the case for holographic theories to all orders in $1/N$. We emphasize that Eq.~\eqref{KWigglyNull} holds purely as a consequence of the validity of the QNEC and the saturation in the vacuum for \(\mR\), two facts which are potentially true in quantum field theories much more generally than free and holographic theories.

Finally, in Sec.~\ref{sec-discussion} we will conclude with a discussion of possible extensions to curved backgrounds and more general regions, connections between the relative entropy and the QNEC, and relations to other work.

\section{Main Argument}\label{sec:main}
\subsection{Review of QNEC}

The von Neumann entropy of a region in quantum field theory can be regarded as a functional of the entangling surface. We will primarily be interested in regions to one side of a cut of a null plane in flat space, for which the entangling surface can be specified by a function \(V(y)\) which indicates the \(v\)-coordinate of the cut as a function of the transverse coordinates, collectively denoted \(y\). See Fig.~\ref{fig-region} for the basic setup. Each cut \(V(y)\) defines a half-space, namely the region to one side of the cut. We will pick the side towards the future of the null plane. For the purposes of this section we are free to consider the more general situation where the entangling surface is only {\em locally} given by a cut of a null plane. Thus the von Neumann entropy can be considered as a functional of a profile $V(y)$ which defines the shape of the entangling surface, at least locally.

Suppose we define a one-parameter family of cuts $V(y;\lambda) \equiv V(y;0) + \lambda \dot V(y)$, with $\dot V(y) >0$ to ensure that $\mathcal{R}(\lambda_1) \subset \mathcal{R}(\lambda_2)$ if $\lambda_1 > \lambda_2$. If $S(\lambda)$ is the entropy of region $\mathcal{R}(\lambda)$, then the QNEC in integrated form states that 
\be\label{nonlocalQNEC}
\int d^{d-2}y \, \braket{T_{vv}(y)} \dot V(y)^2  \geq \frac{\hbar}{2\pi} \frac{d^{2} S}{d\lambda^{2}}.
\ee
In general there would be a \(\sqrt{h}\) induced metric factor weighting the integral, but here and in the rest of the paper we will assume that the \(y\) coordinates have been chosen such that \(\sqrt{h} = 1\). 

By taking advantage of the arbitrariness of $\dot{V}(y)$ we can derive from this the local form of the QNEC. If we take a limit where $\dot{V}(y')^2 \to \delta(y-y')$, then the l.h.s. reduces to $ \braket{T_{vv}}$. We {\em define} $S''(y)$ as the limit of $d^2 S/d\lambda^2$ in the same situation:
\be\label{S''}
\frac{d^{2} S}{d\lambda^{2}} \to S''(y) ~~~~\text{when}~~~~ \dot{V}(y')^2 \to \delta(y-y').
\ee
Taking the limit of the nonlocal QNEC then gives the local one:
\be\label{localQNEC}
\braket{T_{vv}} \geq \frac{\hbar}{2\pi} S''.
\ee
The local QNEC together with strong subadditivity can likewise be used to go backward and derive the nonlocal QNEC~\cite{Bousso:2015mna,Bousso:2015wca,Koeller:2015qmn}. The details of that argument are not important here. In the next section we will discuss the consequences of the saturation of the QNEC, and will have to distinguish whether we mean saturation of the nonlocal inequality Eq.~\eqref{nonlocalQNEC} or the local inequality Eq.~\eqref{localQNEC}, the latter condition being weaker.

\subsection{The QNEC under state perturbations}

In this section we consider how the QNEC behaves under small deformations of the state. We begin with a reference state $\sigma$ and consider the deformed state $\rho = \sigma + \delta \rho$, with $\delta \rho$ traceless but otherwise arbitrary.

Consider a one-parameter family of regions $\mathcal{R}(\lambda)$ as in the previous section. Define $\overline{\mathcal{R}}(\lambda)$ to be the complement of $\mathcal{R}(\lambda)$ within a Cauchy surface. The reduced density operator for any given region $\mathcal{R}(\lambda)$ given by
\be
\rho(\lambda) = \sigma(\lambda) + \delta \rho(\lambda) =  {\rm Tr}_{\overline{\mathcal{R}}(\lambda)} \sigma + {\rm Tr}_{\overline{\mathcal{R}}(\lambda)} \delta \rho.
\ee
By the First Law of entanglement entropy, the entropy of $\rho(\lambda)$ is given by
\be\label{eq-firstlaw}
S(\rho(\lambda)) = S(\sigma(\lambda)) - {\rm Tr}_{\mathcal{R}(\lambda)} \delta \rho(\lambda) \log \sigma(\lambda) + o(\delta\rho^2).
\ee
The second term can be written in a more useful way be defining the modular Hamiltonian $K_\sigma(\lambda)$ as
\be\label{defK}
K_\sigma(\lambda) \equiv - \mathds{1}_{\overline{\mathcal{R}}(\lambda)} \otimes \log \sigma(\lambda).
\ee
Defining $K_\sigma(\lambda)$ this way makes it a global operator, which makes taking derivatives with respect to $\lambda$ formally simpler. Using this definition, we can write Eq.~\eqref{eq-firstlaw} as
\be
S(\rho(\lambda)) = S(\sigma(\lambda)) + {\rm Tr}\,\delta \rho K_\sigma(\lambda) + o(\delta\rho^2).
\ee
Now in the second term the trace is over the global Hilbert space, and the $\lambda$-dependence has been isolated to the operator $K_\sigma(\lambda)$. Taking two derivatives, and simplifying the notation slightly, we find
\be\label{derivativepert}
\frac{d^2S}{d\lambda^2}(\rho) = \frac{d^2S}{d\lambda^2}(\sigma) + {\rm Tr}\,\delta \rho \frac{d^2K_\sigma}{d\lambda^2} + o(\delta\rho^2).
\ee

Suppose that the nonlocal QNEC, Eq.~\eqref{nonlocalQNEC}, is saturated in the state $\sigma$ for all profiles \(\dot{V}(y)\). Then, using Eq.~\eqref{derivativepert}, the nonlocal QNEC for the state $\rho$ can be written as
\begin{align}\label{pertineq}
\int d^{d-2}y \, \left({\rm Tr}\, \delta \rho\, T_{vv}\right) \dot V^2  \geq \frac{\hbar}{2\pi}{\rm Tr}\,\delta \rho \frac{d^2K_\sigma}{d\lambda^2} + o(\delta\rho^2).
\end{align}
The operator $\delta \rho$ was arbitrary, and in particular could be replaced by $-\delta \rho$. Then the only way that Eq.~\ref{pertineq} can hold is if we have the operator equality
\be\label{K''Tkk}
\frac{d^2K_\sigma}{d\lambda^2} = C + \frac{2\pi}{\hbar} \int d^{d-2}y \,  T_{vv} \dot V^2.
\ee
Here $C$ is a number that we cannot fix using this method that is present because of the tracelessness of $\delta\rho$.

Eq.~\eqref{K''Tkk} can be integrated to derive the full modular Hamiltonian \(K_{\sigma}\) if we have appropriate boundary conditions. Up until now we have only made use of local properties of the entangling surface, but in order to provide boundary conditions for the integration of Eq.~\eqref{K''Tkk} we will assume that the entangling surface is {\rm globally} given by a cut of a null plane, and that $V(y;\lambda =0) =0$. We will also make $\sigma$ the vacuum state. In that situation it is known that the QNEC is saturated for free theories, and in the next section we will show that this is also true for holographic theories at all orders in the large-$N$ expansion.

Our first boundary condition is at $\lambda =\infty$.\footnote{It is not always possible to consider the $\lambda\to \infty$ limit of a null perturbation to an entangling surface because parts of the entangling surface may become timelike related to each other at some finite value of $\lambda$, at which point the surface is no longer the boundary of a region on a Cauchy surface. However, when the entangling surface is globally equal to a cut of a null plane this is not an issue.} Since we expect that $K_\sigma(\lambda)$ should have a finite expectation value in any state as $\lambda \to \infty$, it must be that $dK_\sigma/d\lambda\to 0$ as $\lambda\to \infty$. Then integrating Eq.~\eqref{K''Tkk} gives
\be\label{K'Tkk}
\frac{dK_\sigma}{d\lambda} = - \frac{2\pi}{\hbar}  \int d^{d-2}y\int_{V(y;\lambda)}^\infty dv \,  T_{vv} \dot V.
\ee
Note that this equation implies that the vacuum expectation value $\langle K_\sigma(\lambda)\rangle_{\rm vac}$ is actually $\lambda$-independent, which makes vacuum subtraction easy.

Our second boundary condition is Eq.~\eqref{KRindlerNull}, valid at $\lambda =0$ when $V(y;\lambda)=0$. Integrating once morenand making use of this boundary condition, we find 
\be
\Delta K_\sigma(\lambda) = \frac{2\pi}{\hbar} \int d^{d-2}y \int_{V(y;\lambda)}^{\infty} (v-V(y;\lambda)) \,T_{vv} \, dv
\ee
which is Eq.~\eqref{KWigglyNull}. Note that the l.h.s. of this equation is now the vacuum-subtracted modular Hamiltonian.

Before moving on, we will briefly comment on the situation where the local QNEC, Eq.~\eqref{localQNEC}, is saturated but the nonlocal QNEC, Eq.~\eqref{nonlocalQNEC}, is not. Then, analogously to $S''$ in Eq.~\eqref{S''}, one may define a local second derivative of $K_\sigma$:
\be
\frac{d^{2} K_\sigma}{d\lambda^{2}} \to K''_\sigma(y) ~~~~\text{when}~~~~ \dot{V}(y')^2 \to \delta(y-y').
\ee
Very similar manipulations then show that saturation of the local QNEC implies the equality
\be\label{localK''}
K_\sigma'' = \frac{2\pi}{\hbar}  T_{vv}.
\ee
This equation is weaker than Eq.~\eqref{K''Tkk}, which is meant to be true for arbitrary profiles of $\dot{V}(y)$, but it may have a greater regime of validity. We will comment on this further in Sec.~\ref{sec-discussion}.

\section{Holographic Calculation}\label{sec-calculation}

In the previous section we argued that the form of the modular Hamiltonian could be deduced from saturation of the QNEC. In this section we will use the holographic entanglement entropy formula \cite{Ryu:2006bv,Ryu:2006ef,Hubeny:2007xt,Faulkner:2013ana} to show that the QNEC is saturated in vacuum for entangling surfaces defined by arbitrary cuts $v=V(y)$ of the null plane $u=0$ in holographic theories. Our argument applies to any holographic theory defined by a relevant deformation to a holographic CFT, and will be at all orders in the large-$N$ expansion. To reach arbitrary order in $1/N$ we will assume that the all-orders prescription for von Neumann entropy is given by the quantum extremal surface proposal of Engelhardt and Wall~\cite{Engelhardt:2014gca}. This is the same context in which the holographic proof of the QNEC was extended to all orders in $1/N$~\cite{Akers:2016ugt}.\footnote{It is crucial that we demonstrate saturation beyond leading order in large-$N$. The argument in the previous section used {\em exact} saturation, and an error that is na\"ively subleading when evaluated in certain states may become very large in others.}

As before, the entangling surface in the field theory is given by the set of points $\partial \mR =\lbrace (u,v,y): v=V(y), u =0 \rbrace$ with null coordinates $u=t-x$ and $v=t+x$, and the region $\mR$ is chosen to lie in the $u<0$ portion of spacetime. Here $y$ represents $d-2$ transverse coordinates. The bulk quantum extremal surface anchored to this entangling surface is parameterized by the functions $\bar{V}(y,z)$ and $\bar{U}(y,z)$. It was shown in~\cite{Koeller:2015qmn, Akers:2016ugt} that if we let the profile $V(y)$ depend on a deformation parameter $\lambda$, then the second derivative of the entropy is given by
\be\label{S''Uint}
\frac{d^2S}{d\lambda^2} = -\frac{d}{4G\hbar} \int d^{d-2}y \, \frac{d\bar{U}_{(d)}}{d\lambda}~,
\ee
to all orders in $1/N$, where $\bar{U}_{(d)}(y)$ is the coefficient of $z^d$ in the small-$z$ expansion of $\bar{U}(z,y)$. We will show that $\bar{U} =0$ identically for any profile $V(y)$, which then implies that $d^2S/d\lambda^2=0$, which is the statement of QNEC saturation in the vacuum.

One way to show that $\bar{U}$ vanishes is to demonstrate that $\bar{U}=0$ solves the quantum extremal surface equations of motion in the bulk geometry dual to the vacuum state of the boundary theory. The quantum extremal surface is defined by having the sum of the area plus the bulk entropy on one side be stationary with respect to first-order variations of its position. One can show that $\bar{U}=0$ is a solution to the equations of motion if any only if
\be
\frac{\delta S_{\rm bulk}}{\delta \bar{V}(y,z)} = 0
\ee
in the vacuum everywhere along the extremal surface. This would follow from null quantization if the bulk fields were free \cite{Bousso:2015wca}, but that would only allow us to prove the result at order-one in the $1/N$ expansion.

For an all-orders argument, we opt for a more indirect approach using subregion duality, or entanglement wedge reonstruction \cite{Czech:2012bh,Headrick:2014cta,Dong:2016eik,Harlow:2016vwg}.\footnote{The entanglement wedge of a boundary region is the set of bulk points which are spacelike- or null-related to that region's quantum extremal surface on the same side of the quantum extremal surface as the boundary region itself.} A version of this argument first appeared in~\cite{Akers:2016ugt}, and we elaborate on it here.

Entanglement wedge reconstruction requires two important consistency conditions in the form of constraints on the bulk geometry which must hold at all orders in $1/N$: The first constraint, \textit{entanglement wedge nesting} (EWN), states that if one boundary region is contained inside the domain of dependence of another, then the quantum extremal surface associated to the first boundary region must be contained within the entanglement wedge of the second boundary region \cite{Czech:2012bh,Wall:2012uf}. The second constraint, $\mathcal{C} \subseteq \mathcal{E}$, demands that the causal wedge of a boundary region be contained inside the entanglement wedge of that region \cite{Czech:2012bh,Headrick:2014cta,Wall:2012uf,Engelhardt:2014gca,Hubeny:2012wa}. Equivalently, it says that no part of the quantum extremal surface of a given boundary region can be timelike-related to the (boundary) domain of dependence of that boundary region. It was shown in \cite{Akers:2016ugt} that $\mathcal{C} \subseteq \mathcal{E}$ follows from EWN, and EWN itself is simply the statement that a boundary region should contain all of the information about any of its subregions. We will now explain the consequences of these two constraints for $\bar{U}(y,z)$.

Without loss of generality, suppose the region \(\mR\) is defined by a coordinate profile which is positive, $V(y)>0$. Consider a second region $\mR_0$ which has an entangling surface at $v=u=0$ and whose domain of dependence (i.e., Rindler space) contains $\mR$. The quantum extremal surface associated to $\mR_0$ is given by $\bar{U}_0 = \bar{V}_0 =0$. This essentially follows from symmetry.\footnote{One might worry that the quantum extremal surface equations display spontaneous symmetry breaking in the vacuum, but this can be ruled out using $\mathcal{C}\subseteq \mathcal{E}$ with an argument similar to the one we present here.} The entanglement wedge of $\mR_0$ is then a bulk extension of the boundary Rindler space, namely the set of bulk points satisfying $u \leq 0$ and $v \geq 0$. Then EWN implies that $\bar{U} \leq0$ and $\bar{V}\geq 0$.

The only additional constraint we need from $\mathcal{C}\subseteq \mathcal{E}$ is the requirement that the quantum extremal surface for $\mathcal{R}$ not be in the past of the domain of dependence of $\mathcal{R}$. From the definition of $\mathcal{R}$, it is clear that a bulk point is in the past of the domain of dependence of $\mathcal{R}$ if and only if it is in the past of the region $u<0$ on the boundary, which is the same as the region $u<0$ in the bulk. Therefore it must be that $\bar{U} \geq 0$. Combined with the constraint from EWN above, we then conclude that the only possibility is $\bar{U} =0$. This completes the proof that the QNEC is saturated to all orders in $1/N$.

\section{Discussion}\label{sec-discussion}

We conclude by discussing the generality of our analysis, some implications and future directions, and connections with previous work.

\subsection{Generalizations and Future Directions}

\paragraph{General Killing horizons}
Though we restricted to cuts of Rindler horizons in flat space for simplicity, all of our results continue to hold for cuts of bifurcate Killing horizons for QFTs defined in arbitrary spacetimes, assuming the QNEC is true and saturated in the vacuum in this context. In particular, Eq.~\eqref{KWigglyNull} holds with \(v\) a coordinate along the horizon. For holographic theories, entanglement wedge nesting (EWN) and the entanglement wedge being outside of the causal wedge (\(\cse\)) continue to prove saturation of the QNEC. To see this, note that a Killing horizon on the boundary implies a corresponding Killing horizon in the bulk. Now take the reference region \(\mR_{0}\) satisfying \(V(y)=U(y)=0\) to be the boundary bifurcation surface. By symmetry, the associated quantum extremal surface lies on the bifurcation surface of the bulk Killing horizon. Then the quantum extremal surface of the region \(\mR\) defined by \(V(y)\geq0\) must lie in the entanglement wedge of \(\mR_{0}\) --- inside the bulk horizon --- by entanglement wedge nesting, but must also lie on or outside of the bulk horizon by \(\cse\). Thus it lies on the bulk horizon, \(\bar{U}=0\), and the QNEC remains saturated by Eq.~\eqref{S''Uint}.

\paragraph{Future work}
In this work, we have only established the form of $K_{\mathcal{R}}$ for regions $\mathcal{R}$ bounded by arbitrary cuts of a null plane. A natural next direction would be to understand if and how we can extend Eq.~\eqref{localK''} to more general entangling surfaces. As discussed above, the QNEC was shown to hold for locally flat entangling surfaces in holographic, free and super-renormalizable field theories \cite{Bousso:2015wca, Koeller:2015qmn, Akers:2016ugt}. Thus, if we could prove saturation, i.e. that $S_{\rm vac}^{\prime \prime}=0$ at all orders in $1/N$, then we would establish \eqref{localK''} for all regions with a locally flat boundary.

One technique to probe this question is to perturb the entangling surface away from a flat cut and compute the contributions to the QNEC order-by-order in a perturbation parameter \(\epsilon\). Preliminary calculations \cite{futurework} have revealed that for holographic theories at leading order in large \(N\), $S_{\rm vac}^{\prime \prime} = 0$ at all orders in \(\epsilon\). 

Another interesting problem is to show that in a {\it general} QFT vacuum, null derivatives of entanglement entropy across arbitrary cuts of null planes vanish.
That, along with a general proof of QNEC will establish (18) as a consequence. We will leave this to future work.

\subsection{The QNEC as \(S(\rho\|\sigma)'' \geq 0\)}\label{sec:SrelQNEC}
There is a connection between the QNEC and relative entropy, first pointed out in \cite{Akers:2016ugt}, that we elaborate on here. The relative entropy \(S(\rho\|\sigma)\) between two states \(\rho\) and \(\sigma\) is defined as
\begin{align}\label{SrelDef}
	S(\rho\|\sigma) = \Tr{\rho \log{\rho}} - \Tr{ \rho \log{\sigma}}
\end{align}
and provides a measure of distinguishability between the two states \cite{Nielsen:2011:QCQ:1972505}. Substituting the definition of \(K\), Eq.~\eqref{Kdef}, into Eq.~\eqref{SrelDef} provides a useful alternate presentation:
\begin{align}
	S(\rho\|\sigma) = \langle K_\sigma\rangle_\rho - S(\rho).
\end{align}
If Eq.~\eqref{KWigglyNull} is valid, then taking two derivatives with respect to a deformation parameter, as in the main text, shows that the nonlocal QNEC, Eq.~\eqref{nonlocalQNEC}, is equivalent to 
\begin{align}\label{eq-srelqnec}
	\partial_\lambda^2 S(\rho(\lambda) \|\sigma(\lambda)) \geq 0.
\end{align}
For comparison, monotonicity of relative entropy for the types of regions and deformations we have been discussing can be written as
\begin{align}
	\partial_\lambda S(\rho(\lambda) \|\sigma(\lambda)) \leq 0.
\end{align}
Eq.~\eqref{eq-srelqnec} is a sort of ``convexity" of relative entropy.\footnote{This is distinct from the well-known convexity of relative entropy, which says that $S(t\rho_1 + (1-t)\rho_2 \| \sigma) \leq tS(\rho_1\| \sigma)  + (1-t)S(\rho_2 \| \sigma)$.} Unlike monotonicity of relative entropy, which says that the first derivative is non-positive, there is no general information-theoretic reason for the {\em second} derivative to be non-negative. In the event that Eq.~\eqref{localK''} holds but not Eq.~\eqref{KWigglyNull}, we would still have 
\be
S(\rho \| \sigma)'' \geq 0.
\ee
where the $''$ notation denotes a local deformation as in Sec.~\ref{sec:main}.

It would be extremely interesting to characterize what about quantum field theory and null planes makes \eqref{eq-srelqnec} true. We can model the null deformation as a non-unitary time evolution in the space of states, with the vacuum state serving as an equilibrium state for this evolution. Then an arbitrary finite-energy state will relax toward the equilibrium state, with the relative entropy $S(\rho\|\sigma)$ characterizing the free energy as a function of time. Monotonicity of relative entropy is then nothing more than the statement that free energy decreases, i.e. the second law of thermodynamics. The second derivative statement gives more information about the approach to equilibrium. If that approach is of the form of exponential decay, then all successive derivatives would alternate in sign. However, for null deformations in quantum field theory we do not expect to have a general bound on the behavior of derivatives of the energy-momentum tensor, meaning that the third derivative of the free energy should not have a definite sign.\footnote{We thank Aron Wall for a discussion of this point.}  Perhaps there is some way of characterizing the approach to equilibrium we have here, which is in some sense smoother than the most general possibility but not so constrained as to force exponential behavior.

\subsection{Relation to previous work}
Faulkner, Leigh, Parrikar and Wang \cite{Faulkner:2016mzt} have discussed results very similar to the ones presented here. They demonstrated that for first-order null deformations \(\delta V(y)\) to a flat cut of a null plane, the perturbation to the modular Hamiltonian takes the form 
\begin{equation}\label{modhamfirstorder}
\braket{K_{\mR}}_{\psi} - \braket{K_{\mR_0}}_{\psi} = -\frac{2\pi}{\hbar} \int d^{d-2}y \int_{V(y)}  dv \,T_{vv}(y)\, \delta V(y)
\end{equation}
This is precisely the form expected from our equation (\ref{KWigglyNull}). Faulkner et al. went on to suggest that the natural generalization of the modular Hamiltonian to finite deformations away from a flat cut takes the form of Eq.~\eqref{KWigglyNull}. In the context of holography they showed that this conclusion applied both on the boundary and in the bulk is consistent with JLMS \cite{Jafferis:2015del}. In the present paper, we have shown that Eq.~\eqref{KWigglyNull} holds for theories which obey the QNEC, and for which the QNEC is saturated in the vacuum. A non-perturbative, field theoretic proof of these assumptions remains a primary goal of future work.


\acknowledgments
We would like to thank Chris~Akers, Raphael~Bousso, Aron~Wall and Zach~Fisher for discussions and correspondence. This work was supported in part by the Berkeley Center for Theoretical Physics, by the National Science Foundation (award numbers 1521446, and 1316783), by FQXi, and by the US Department of Energy under contract DE-AC02-05CH11231.

\bibliographystyle{utcaps}
\bibliography{all}
\end{document}